\documentclass[a4paper,10pt]{article}
\pdfoutput=1
\usepackage[T1]{fontenc}
\usepackage{eurosym}
\usepackage{lastpage}
\usepackage{amsfonts}
\usepackage{bbm}
\usepackage{cite}
\usepackage{xspace}
\usepackage[margin=20mm,includehead,includefoot]{geometry}
\usepackage{fancyhdr}
\usepackage{booktabs}
\usepackage{graphicx}
\usepackage{multirow}
\usepackage{array}
\usepackage{xcolor}
\usepackage{csquotes}
\usepackage{pgfgantt}
\usepackage{slashed}
\usepackage{rotating}
\usepackage[graphicx]{realboxes}
\usepackage{hyperref}
\usepackage{amsmath,amssymb}
\usepackage[normalem]{ulem}
\bibliography{bibliography}

\newcolumntype{M}[1]{>{\centering\arraybackslash}m{#1}}
\newcolumntype{N}{@{}m{0pt}@{}}
\newcommand{\be}{\begin{equation}}
\newcommand{\ee}{\end{equation}}
\newcommand{\ba}{\begin{array}}
\newcommand{\ea}{\end{array}}
\newcommand{\bea}{\begin{eqnarray}}
\newcommand{\eea}{\end{eqnarray}}
\newcommand{\p}{\partial}

\newcommand{\nn}{\nonumber}

\newcommand{\g}{\gamma}
\newcommand{\m}{\mu}

\usepackage{pgfgantt}
\usepackage{xcolor}
\usepackage[utf8]{inputenc}

\definecolor{barblue}{RGB}{153,204,254}
\definecolor{groupblue}{RGB}{51,102,254}
\definecolor{linkred}{RGB}{165,0,33} 

\usepackage{xcolor,colortbl}
\begin{document}
\baselineskip 24pt

\begin{center}
{\Large \bf Non-relativistic Conformal Field Theory in Momentum Space}

\end{center}

\vskip .6cm
\medskip

\vspace*{4.0ex}

\baselineskip=18pt

\centerline{\large \rm  Rajesh Kumar Gupta, Meenu}

\vspace*{4.0ex}

\centerline{\large \it Department of Physics, Indian Institute of Technology Ropar, }

\centerline{\large \it  Rupnagar, Punjab 140001, India}

\vspace*{1.0ex}
\centerline{E-mail: rajesh.gupta@iitrpr.ac.in, meenu.20phz0003@iitrpr.ac.in }

\vspace*{5.0ex}

\centerline{\bf Abstract} \bigskip
Non-relativistic conformal field theory describes many-body physics at unitarity. The correlation functions of the system are fixed by the requirement of the conformal invariance.
In this article, we discuss the correlation functions of scalar operators in non-relativistic conformal field theories in momentum space. We discuss the solution of conformal Ward identities and express 2,3, and 4-point functions as a function of energy and momentum. We also express the 3- and 4-point functions as the one-loop and three-loop Feynman diagram computations in the momentum space. Lastly, we generalize the discussion to the momentum space correlation functions in the presence of a boundary. 
\vfill \eject

\baselineskip=18pt

%\tableofcontents
\section{Introduction}
The critical point of non-relativistic many-body systems is described by a scale-invariant non-relativistic field theory. Such a theory is invariant under the Galilean transformation together with the anisotropic scale transformation. A special case of scale invariant systems with anisotropic exponent $z=2$ is the system described by a non-relativistic conformal field theory (NRCFT). The space-times symmetry group of such theory is the Schr$\ddot{\text{o}}$dinger group, which consists of an expansion transformation in addition to the scale and Galilean transformations. The operators in NRCFT are also labelled by a particle number, which appears in the conformal algebra as the central extension of the Galilean algebra. For the details of NRCFT and its application in many body systems, see~\cite{Hagen:1972pd, Henkel:1993sg, Mehen:1999nd, Nishida:2007pj, Nishida:2006br, Nishida:2006eu, Braaten:2008uh, Golkar:2014mwa, Goldberger:2014hca, Gupta:2022azd, Gupta:2022mer}.

Conformal invariance restricts the form of the correlation function. In this article, we will confine ourselves to scalar operators. On the scalar operators, the conformal invariance requires the correlation function to satisfy the following Ward identities.
\bea\label{WardidSpatial}
&&\sum_{a=1}^{n}\Big(2\tau_{a}\frac{\p}{\p \tau_{a}}+x_{ai}\frac{\p}{\p x_{ai}}+\Delta_{a}\Big)<{\mathcal T}O_{1}(x_{1})......O_{n}(x_{n})>=0\,,\nn\\
&&\sum_{a=1}^{n}\Big(\tau_{a}^{2}\frac{\p}{\p \tau_{a}}+\tau_{a}x_{ai}\frac{\p}{\p x_{ai}}+\tau_{a}\Delta_{a}+\frac{N_{a}}{2}x_{a}^{2}\Big)<{\mathcal T}O_{1}(x_{1})......O_{n}(x_{n})>=0\,,\nn\\
&&\sum_{a=1}^{n}\Big(\tau_{a}\frac{\p}{\p x_{ai}}+N_{a}x_{ai}\Big)<{\mathcal T}O_{1}(x_{1})......O_{n}(x_{n})>=0\,.
\eea
The first equation follows from the dilatation, the second from the expansion transformation and the last from the boost transformation. Here $<{\mathcal T}O_{1}(x_{1})......O_{n}(x_{n})>$ is the time-ordered product of scalar operators of particle number $N_{a}$ and scaling dimensions $\Delta_{a}$, and $x_{a}=(\tau_{a},\vec x_{a})$. 
On top of these, we also have the condition of space and time translational invariance. In the above, we are working in the Euclidean picture where we have replaced the conventional $t$ by $i\tau$. 

A general structure of the correlation function in the position space can be obtained by solving the above Ward identities. The correlation functions also need to satisfy the particle number conservation, i.e. a non-zero correlation function exists, provided
\be
\sum_{a=1}^{n}N_{a}=0\,.
\ee
Solving Ward identities for 2, 3 and 4-point functions, we obtain (taking $\tau_{1}>\tau_{2}>....>\tau_{n}$)
\bea\label{FourierTrnsf}
&&<O_{1}(x_{1})O_{2}(x_{2})>=\delta_{\Delta_{1},\Delta_{2}}\frac{C_{\mathcal O}}{\tau_{12}^{\Delta}}e^{-\frac{N_{1}\vec x_{12}^{2}}{2\tau_{12}}}\,,\nn\\
&&<O_{1}(x_{1})O_{2}(x_{2})O_{3}(x_{3})>=e^{-\frac{N_{1}}{2}\frac{\vec x_{13}^{2}}{\tau_{13}}-\frac{N_{2}}{2}\frac{\vec x_{23}^{2}}{\tau_{23}}}\tau_{12}^{\Delta_{12}}\tau_{13}^{\Delta_{13}}\tau_{23}^{\Delta_{23}}f(v_{123})\,,\\
&&<O_{1}(x_{1})O_{2}(x_{2})O_{3}(x_{3})O_{4}(x_{4})>=e^{-\frac{N_{1}}{2}\frac{\vec x_{14}^{2}}{\tau_{14}}-\frac{N_{2}}{2}\frac{\vec x_{24}^{2}}{\tau_{24}}-\frac{N_{3}}{2}\frac{\vec x_{34}^{2}}{\tau_{34}}}\prod_{i<j}\tau_{ij}^{\frac{\Delta}{6}-\frac{\Delta_{i}+\Delta_{j}}{2}}h(v_{124},v_{134},v_{234},\frac{\tau_{12}\tau_{34}}{\tau_{13}\tau_{24}})\,.\nn
\eea
In the above $\tau_{ij}=\tau_{i}-\tau_{j}$ and $\vec x_{ij}=\vec x_{i}-\vec x_{j}$ and $\Delta=\sum_{a=1}^{n}\Delta_{a}$. Also, the combinations of scaling dimensions that appear in the 3-point function are
\be
\Delta_{12}=\frac{\Delta}{2}-(\Delta_{1}+\Delta_{2}),\quad \Delta_{13}=\frac{\Delta}{2}-(\Delta_{1}+\Delta_{3}),\quad \Delta_{23}=\frac{\Delta}{2}-(\Delta_{3}+\Delta_{2})\,.
\ee
Finally, the conformal cross ratios that appear in the above are 
\be
v_{ijk}=\frac{\vec x_{jk}^{2}}{\tau_{jk}}+\frac{\vec x_{ij}^{2}}{\tau_{ij}}-\frac{\vec x_{ik}^{2}}{\tau_{ik}}\,,\quad i<j<k\,.
\ee

Quantum field theory in the position space is useful to reveal various properties of the underlying system, such as correlation length, causality, OPE, etc. However, generically, it is convenient to perform the computation of scattering amplitude and Green's functions using Feynman diagrams in the momentum space.  Algebras in these computations are simpler in the momentum space than in the position space. This motivates to study the conformal field theory in the momentum space. Several works have been done in understanding the implications of conformal Ward identities in the momentum space, see~\cite{Bzowski:2013sza, Bzowski:2015pba, Bzowski:2019kwd}. 
In the present article, we would like to carry out a similar analysis in the NRCFT context, i.e. finding the general form of the correlation function in the energy-momentum space. We will find the expression for 2-,3- and 4-point functions in momentum space. We will see that the correlation function may not be well-defined for all scaling dimensions. For example, the 2-point function is divergent for $\Delta=\frac{d}{2}+1+n$, where $n$ is a non-negative integer.
Making sense of the correlation function requires regularization and renormalization. We will discuss this in the case of the 2-point function.

{\bf Note added:} While this draft was in preparation, an article~\cite{S:2024zqp} appeared on the arXiv. The article presents the conformally invariant quantum mechanics in momentum space. We find that, even though the contexts are different, many of our results have similarities with the results obtained in the paper. 
%%%%%%%%%%%%%%%%%%%%%%%%%%%%%%%%%%%%%%%%%%%%%%%%
\section{Solving Conformal Ward identities in Momentum Space}
Let us consider the following Fourier transformations of a scalar operator
\be
O(\tau,\vec x)=\int\frac{dE\,d^{d}p}{(2\pi)^{d+1}}e^{iE\tau+i\vec p\cdot \vec x}O(E,\vec p),\quad O(E,\vec p)=\int\,d\tau\,d^{dx}e^{-iE\tau-i\vec p\cdot \vec x}O(\tau,\vec x)\,.
\ee
Then time and space translation invariance implies that
\be
<O_{1}(E_{1},\vec p_{1}).....O_{n}(E_{n},\vec p_{n})>=(2\pi)^{d+1}\delta(E_{1}+...+E_{n})\delta^{d}(\vec p_{1}+...+\vec p_{n})<<O_{1}(E_{1},\vec p_{1})......O_{n}(E_{n},\vec p_{n})>>\,.
\ee
The dilatation Ward identity in the momentum space becomes
\be
\Big[\sum_{a=1}^{n}\Delta_{a}-(n-1)(d+2)-\sum_{a=1}^{n-1}\Big(p_{a,i}\frac{\p}{\p p_{a,i}}+2E_{a}\frac{\p}{\p E_{a}}\Big)\Big]<<O_{1}(E_{1},\vec p_{1})......O_{n}(E_{n},\vec p_{n})>>=0\,.
\ee
The expansion Ward identity is
\be
\sum_{a=1}^{n-1}\Big(-i(d+2-\Delta_{a})\frac{\p}{\p E_{a}}-iE_{a}\frac{\p^{2}}{\p E_{a}^{2}}-ip_{ai}\frac{\p^{2}}{\p E_{a}\p p_{ai}}-\frac{N_{a}}{2}\frac{\p^{2}}{\p p_{ai}\p p_{ai}}\Big)<<O_{1}(E_{1},\vec p_{1})......O_{n}(E_{n},\vec p_{n})>>=0\,.
\ee
Finally, the boost Ward identity is
\be
\sum_{a=1}^{n-1}\Big(-p_{ai}\frac{\p}{\p E_{a}}+iN_{a}\frac{\p}{\p p_{ai}}\Big)<<O_{1}(E_{1},\vec p_{1})......O_{n}(E_{n},\vec p_{n})>>=0\,.
\ee
{\bf 2-point function:} We start with the 2-point function $<<O_{1}(E,\vec p)O_{2}(-E,-\vec p)>>$. The dilatation Ward identity gives the differential equation
\be
\Big[(\Delta-(d+2))-\Big(p_{i}\frac{\p}{\p p_{i}}+2E\frac{\p}{\p E}\Big)\Big]<<O_{1}(E,\vec p)O_{2}(-E,-\vec p)>>=0\,.
\ee
Here $\Delta=\Delta_{1}+\Delta_{2}$.
This implies that
\be
<<O_{1}(E,\vec p)O_{2}(-E,-\vec p)>>=E^{\frac{\Delta-(d+2)}{2}}g(\frac{p^{2}}{E})\,.
\ee
Next, we look at the boost Ward identity. We get
\be
(-p_{i}\frac{\p}{\p E}+iN_{1}\frac{\p}{\p p_{i}})(E^{\frac{\Delta-(d+2)}{2}}g(\frac{p^{2}}{E}))=0\,,
\ee
which simplifies to
\be
-\Big(\frac{\Delta-(d+2)}{2}g(\frac{p^{2}}{E})-g'(\frac{p^{2}}{E})\frac{p^{2}}{E}\Big)+2iN_{1}g'(\frac{p^{2}}{E})=0\,.
\ee
The solution to the above equation is
\be
g(\frac{p^{2}}{E})=(\frac{p^{2}}{E}+2iN_{1})^{\frac{\Delta-(d+2)}{2}}\,.
\ee
Thus, the two point function becomes
\be\label{2PtFn}
<<O_{1}(E,\vec p)O_{2}(-E,-\vec p)>>=c'\,\Big(\frac{p^{2}}{2N_{1}}+iE\Big)^{\frac{\Delta-(d+2)}{2}}\,,
\ee
where $c'$ is a constant.\\
Next, we see the Ward identity for the expansion transformation. The differential equation is
\be
\Big(-2i\frac{\p}{\p E}-iE\frac{\p^{2}}{\p E^{2}}-id\frac{\p}{\p E}-ip_{i}\frac{\p^{2}}{\p E\p p_{i}}+i\Delta_{1}\frac{\p}{\p E}-\frac{N_{1}}{2}\frac{\p^{2}}{\p p_{i}\p p_{i}}\Big)<<O_{1}(E,\vec p)O_{2}(-E,-\vec p)>>=0\,.
\ee
The 2-point function~\eqref{2PtFn} solves the above provided the following condition is satisfied,
\be
\Delta_{1}=\Delta_{2}\,.
\ee
Thus, the 2-point function is
\be
<<O_{1}(E,\vec p)O_{2}(-E,-\vec p)>>=c'\,\delta_{\Delta_{1},\Delta_{2}}\Big(\frac{p^{2}}{2N_{1}}+iE\Big)^{\Delta_{1}-\frac{d}{2}-1}\,.
\ee
In a non-relativistic conformal field theory, the scaling dimension of an operator satisfies the unitarity bound. All operators have the scaling dimension $\Delta_{a}\geq \frac{d}{2}$. Looking at the above 2-point function, we observe that the 2-point function of the operators with scaling dimensions $\Delta_{n}=\frac{d}{2}+1+n$, with $n=0,1,2,...$, are local. This is a situation very similar to the relativistic CFTs in momentum space. In the position space, this would correspond to the following 2-point function, 
\be
<O_{1}(\tau,\vec x)O^{\dagger}_{1}(0,\vec 0)>=\Big(\p_{\tau}-\frac{\nabla^{2}}{2N_{1}}\Big)^{n}\delta(\tau)\delta^{d}(\vec x)\,.
\ee
As a consequence, the correlation function can be set to zero by adding the local counter term of the form
\be
J(\tau,\vec x)^{\dagger}\Big(\p_{\tau}-\frac{\nabla^{2}}{2N_{1}}\Big)^{n}J(\tau,\vec x)\,,
\ee
where $J(\tau,\vec x)$ is the source of the operator $O_{1}(\tau,\vec x)$. This would imply that $O_{1}(\tau,\vec x)$ must be a null operator. Generically, this is not the case. For example, if we consider $d=2$ and $\phi^{\dagger}$ is the scalar operator with scaling dimension $\Delta_{\phi}=1$ and particle number $+1$, then we see that $\phi^{\dagger\,r}$, with $r\geq 2$, is a non-trivial operator with scaling dimension $\Delta_{r}=r$ and particle number $r$. 
A related issue which we will discover in the next section while performing the Fourier transform is that the transform exists provided $\Delta_{a}\neq\frac{d}{2}+1+n$.

The regularization of the 2-point function for the operator of the dimension $\Delta_{1}=\frac{d}{2}+1+n$ can be performed in a very similar manner as done in~\cite{Bzowski:2015pba}. We work in the dimensional regularization where~\footnote{In the non-relativistic case, where space and time are not at equal footing, there are different possible ways of doing dimensional regularization, e.g. one can regularize both space and time dimensions separately. Furthermore, as noted in the paper~\cite{Bzowski:2015pba}, this regularization may lead to a new kind of anomaly in the non-relativistic systems~\cite{Gupta2024}.}
\be
d\rightarrow d'=2+2\kappa_{1}\epsilon,\quad \Delta\rightarrow\Delta_{1}'=\Delta_{1}+(\kappa_{1}+\kappa_{2})\epsilon
\ee
In this case, the dilatation ($\mathcal D$) and expansion ($\mathcal C$) differential operators are modified as
\be
\tilde{\mathcal D}=\mathcal D+2\kappa_{2}\epsilon,\quad \tilde{\mathcal C}=\mathcal C-i(\kappa_{1}-\kappa_{2})\epsilon\frac{\p}{\p E}-{\frac{N_{1}\kappa_{1}\epsilon}{p}\frac{\p}{\p p}}\,.
\ee
Then the two point function in the dimensions $d'+1$ is
\bea
<<O_{1}(E,\vec p)O_{2}(-E,-\vec p)>>_{\text{regul}}&=&c(\kappa_{1},\kappa_{2},\epsilon)\,\Big(\frac{p^{2}}{2N_{1}}+iE\Big)^{\Delta_{1}'-\frac{d'}{2}-1}\nn\\
&=&c(\kappa_{1},\kappa_{2},\epsilon)\,\Big(\frac{p^{2}}{2N_{1}}+iE\Big)^{\Delta_{1}-\frac{d}{2}-1}\Big(\frac{p^{2}}{2N_{1}}+iE\Big)^{\kappa_{2}\epsilon}\,.
\eea
In the regularized dimensions, the coefficient $c(\kappa_{1},\kappa_{2},\epsilon)$ will have simple pole in $\epsilon$, i.e.
\be
c(\kappa_{1},\kappa_{2},\epsilon)=\frac{c_{-1}}{\epsilon}+c_{0}+\mathcal O(\epsilon)\,.
\ee 
Then the regularised two point function is
\be
<<O_{1}(E,\vec p)O_{2}(-E,-\vec p)>>_{\text{regul}}=\Big(\frac{p^{2}}{2N_{1}}+iE\Big)^n\Big[\frac{c_{-1}}{\epsilon}+c_{-1}\kappa_{2}\ln(\frac{p^{2}}{2N_{1}}+iE)\Big]\,.
\ee
The divergent term is local and can be removed by the counter term 
\be
\int d\tau\,d^{d+2\kappa_{1}\epsilon}x\,\mu^{\kappa_{2}\epsilon}J^{\dagger}\Big(\p_{\tau}-\frac{\nabla^{2}}{2N_{1}}\Big)^{n}J\,.
\ee
The counter term introduces a scale $\mu$ and the renormalized two point function becomes
\be\label{renorm2PtFn}
<<O_{1}(E,\vec p)O_{2}(-E,-\vec p)>>_{\text{renorm.}}=c\,\Big(\frac{p^{2}}{2N_{1}}+iE\Big)^n\ln\frac{(\frac{p^{2}}{2N_{1}}+iE)}{\m}\,.
\ee
{\bf 3-point function:} Next, we discuss the 3-point function. The Ward identities are as follows: the dilatation Ward identity is
\be
\Big[(\Delta-2(d+2))-\Big(p_{1,i}\frac{\p}{\p p_{1,i}}+p_{2,i}\frac{\p}{\p p_{2,i}}+2E_{1}\frac{\p}{\p E_{1}}+2E_{2}\frac{\p}{\p E_{2}}\Big)\Big]f(E_{1},E_{2},\vec p_{1},\vec p_{2})=0\,.
\ee
Here $\Delta=\Delta_{1}+\Delta_{2}+\Delta_{3}$. The boost Ward identity is
\be
\Big[i(N_{1}\frac{\p}{\p p_{1,i}}+N_{2}\frac{\p}{\p p_{2,i}})-(p_{1,i}\frac{\p}{\p E_{1}}+p_{2,i}\frac{\p}{\p E_{2}})\Big]f(E_{1},E_{2},\vec p_{1},\vec p_{2})=0\,.
\ee
and the expansion Ward identity becomes
\bea
&&\Big(-i(d+2-\Delta_{1})\frac{\p}{\p E_{1}}-i(d+2-\Delta_{2})\frac{\p}{\p E_{2}}-iE_{1}\frac{\p^{2}}{\p E_{1}^{2}}-iE_{2}\frac{\p^{2}}{\p E_{2}^{2}}\nn\\
&&-ip_{1i}\frac{\p^{2}}{\p E_{1}\p p_{1i}}-ip_{2i}\frac{\p^{2}}{\p E_{2}\p p_{2i}}-\frac{N_{1}}{2}\frac{\p^{2}}{\p p_{1i}\p p_{1i}}-\frac{N_{2}}{2}\frac{\p^{2}}{\p p_{2i}\p p_{2i}}\Big)f(E_{1},E_{2},\vec p_{1},\vec p_{2})=0\,.
\eea
Here $f(E_{1},E_{2},\vec p_{1},\vec p_{2})=<<O_{1}(E_{1},\vec p_{1})O_{2}(E_{2},\vec p_{2})O_{3}(E_{3},\vec p_{3})>>$.

Next, we want to obtain the general solution of the above differential equations. It is clear that the rotation invariance implies that $f(E_{1},E_{2},\vec p_{1},\vec p_{2})$ will be a function of $E_{1,2}$ and $\vec p^{2}_{1},\,\vec p^{2}_{2},\,\vec p_{3}^{2}$. It is convenient to work in the variables $(z_{1},z_{2},p_{1},p_{2},p_{3})$ where
\be
iz_{1}=iE_{1}+\frac{\vec p_{1}^{2}}{2N_{1}},\quad iz_{2}=iE_{2}+\frac{\vec p_{2}^{2}}{2N_{2}}\,.
\ee
In terms of these variables, the boost Ward identity becomes the following set of equations
\bea
&&\Big(\frac{N_{1}}{p_{1}}\frac{\p}{\p p_{1}}-\frac{N_{3}}{p_{3}}\frac{\p}{\p p_{3}}\Big)f(z_{1},z_{2},p_{1},p_{2},p_{3})=0\,,\nn\\
&&\Big(\frac{N_{2}}{p_{2}}\frac{\p}{\p p_{2}}-\frac{N_{3}}{p_{3}}\frac{\p}{\p p_{3}}\Big)f(z_{1},z_{2},p_{1},p_{2},p_{3})=0\,,\nn\\
&&\Big(\frac{N_{1}}{p_{1}}\frac{\p}{\p p_{1}}-\frac{N_{2}}{p_{2}}\frac{\p}{\p p_{2}}\Big)f(z_{1},z_{2},p_{1},p_{2},p_{3})=0\,.
\eea
Clearly not all three equations are independent. The above implies that the function $f(z_{1},z_{2},p_{1},p_{2},p_{3})$ depends on the variable
\be
q=\frac{p_{1}^{2}}{2N_{1}}+\frac{p_{2}^{2}}{2N_{2}}+\frac{p_{3}^{2}}{2N_{3}}\,.
\ee
Thus, we have the function $f(z_{1},z_{2},q)$. Furthermore, the dilatation Ward identity implies that
\be
f(z_{1},z_{2},q)=z_{1}^{\frac{\Delta}{2}-d-2}\tilde f(u,v)\,,
\ee
where $u=-\frac{z_{2}}{z_{1}}$ and $v=-i\frac{q}{z_{1}}$. 

Next, we look at the expansion Ward identity in $z_{i}$ and $q$ coordinates. We get
\be
\Big(-i(\frac{d}{2}+2-\Delta_{1})\frac{\p}{\p z_{1}}-i(\frac{d}{2}+2-\Delta_{2})\frac{\p}{\p z_{2}}-iz_{1}\frac{\p^{2}}{\p z_{1}^{2}}-iz_{2}\frac{\p^{2}}{\p z_{2}^{2}}-\frac{d}{2}\frac{\p}{\p q}-q\frac{\p^{2}}{\p q^{2}}\Big)f(z_{1},z_{2},q)=0\,.
\ee
In the variables $(u,v)$, the differential equation becomes
\bea\label{MasterEq}
&&u(1-u)\frac{\p^{2}\tilde f}{\p u^{2}}+v(1-v)\frac{\p^{2}\tilde f}{\p v^{2}}-2uv\frac{\p^{2}\tilde f}{\p u\p v}+(-\frac{3d}{2}-\Delta_{1}+\Delta-4)\Big(u\frac{\p\tilde f}{\p u}+v\frac{\p\tilde f}{\p v}\Big)\nn\\
&&+(\frac{d}{2}+2-\Delta_{2})\frac{\p\tilde f}{\p u}+\frac{d}{2}\frac{\p\tilde f}{\p v}-(\frac{\Delta}{2}-d-2)(-\frac{d}{2}-\Delta_{1}+\frac{\Delta}{2}-1)\tilde f=0\,.
\eea
Note that using the condition $E_{1}+E_{2}+E_{3}=0$, we could also write $v$ as
\be
v=1-u-w,\quad\text{where}\quad w=-\frac{iE_{3}+\frac{p_{3}^{2}}{2N_{3}}}{iE_{1}+\frac{p_{1}^{2}}{2N_{1}}}\,.
\ee
Let us now discuss the solution of the differential equation~\eqref{MasterEq}.
Note that it is a differential equation in two variables, and because of the term $uv\frac{\p^{2}\tilde f}{\p u\p v}$, we can not have the solution using the separation of variables. Let us look for a solution in the following form
\be\label{Ansatz1}
\tilde f(u,v)=\sum_{n=0}^{\infty}v^{n}C_{n}(u)\,.
\ee
One can also look for the solution as a Taylor series expansion in powers of $u$. This may provide a set of bootstrap equations in the NRCFT.

Substituting the ansatz~\eqref{Ansatz1} in~\eqref{MasterEq}, we obtain
\bea
&&\sum_{n}\Big(u(1-u)v^{n}\frac{\p^{2}C_{n}}{\p u^{2}}+(1-v)n(n-1)v^{n-1}C_{n}-2unv^{n}\frac{\p C_{n}}{\p u}+v^{n}(u\frac{\p C_{n}}{\p u}+nC_{n})(-\frac{3d}{2}-\Delta_{1}+\Delta-4)\nn\\
&&+(\frac{d}{2}+2-\Delta_{2})v^{n}\frac{\p C_{n}}{\p u}+\frac{d}{2}n v^{n-1}C_{n}-(\frac{\Delta}{2}-d-2)(-\frac{d}{2}-\Delta_{1}+\frac{\Delta}{2}-1)v^{n}C_{n}\Big)=0\,.
\eea
Then, comparing the coefficient of $v^{n}$, we obtain
\be\label{HypergeometricEq}
u(1-u)\frac{\p^{2}C_{n}}{\p u^{2}}+[c-(a_{n}+b_{n}+1)u]\frac{\p C_{n}}{\p u}-a_{n}b_{n} C_{n}=-(n+1)(n+\frac{d}{2})C_{n+1}\,.
\ee
Here
\be
c=\frac{d}{2}+2-\Delta_{2}\,,
\ee
and $a_{n}$ and $b_{n}$ are solutions of equations
\be
a_{n}+b_{n}=2n+\frac{3d}{2}+\Delta_{1}-\Delta+3,\quad a_{n}b_{n}=n(n-1)+n(\frac{3d}{2}+\Delta_{1}-\Delta+4)-(\frac{\Delta}{2}-d-2)(\frac{d}{2}+\Delta_{1}-\frac{\Delta}{2}+1)\,.
\ee
In particular, we note a solution for $a_{n}$ and $b_{n}$ as given by
\be
a_{n}=2+d+n-\frac{\Delta}{2},\quad b_{n}=1+\frac{d}{2}+n-\frac{\Delta}{2}+\Delta_{1}\,.
\ee
The differential equation \eqref{HypergeometricEq} can be written as
\be
\mathcal D^{(n)}C_{n}=-(n+1)(n+\frac{d}{2})C_{n+1}\,.
\ee
The above tells us that if we know say $C_{0}(u)$, then we can determine $C_{1}(u)$ and using the recursion relations, we can determine all the coefficients $C_{n}(u)$, 
\be
\mathcal D^{(n-1)}\mathcal D^{(n-2)}......\mathcal D^{(0)}C_{0}=4!(-1)^{n}\Big(\frac{d}{2}\Big)_{n}C_{n}\,,
\ee
where $(q)_{n}=q(1+q)....(n-1+q)$.
Thus, the complete solution is
\be
\tilde f(u,v)=\sum_{n=0}^{\infty}v^{n}\frac{(-1)^{n}}{4!\Big(\frac{d}{2}\Big)_{n}}\mathcal D^{(n-1)}\mathcal D^{(n-2)}......\mathcal D^{(0)}C_{0}(u)\,.
\ee
We see that the solution is undetermined up to a function of $u$, i.e. $C_{0}(u)$. This is a reflection of the fact that the 3-point function is determined up to a function of a cross-ratio. Thus, the 3-point function is
\be\label{3PtFn2}
<<O_{1}(E_{1},\vec p_{1})O_{2}(E_{2},\vec p_{2})O_{3}(E_{3},\vec p_{3})>>=\Big(iE_{1}+\frac{\vec p_{1}^{2}}{2N_{1}}\Big)^{\frac{\Delta}{2}-d-2}\sum_{n=0}^{\infty}v^{n}\frac{(-1)^{n}}{4!\Big(\frac{d}{2}\Big)_{n}}\mathcal D^{(n-1)}\mathcal D^{(n-2)}......\mathcal D^{(0)}C_{0}(u)\,.
\ee

We note here an interesting connection with the Appell $F_{2}$ hypergeometric function, see for example~\cite{Ananthanarayan:2021bqz} and references therein. The Appell $F_{2}$ hypergeometric function is a solution of the following coupled differential equations in two variables $x$ and $y$:   
\bea\label{AppellF2}
&&x(1-x)\frac{\p^{2}g(x,y)}{\p x^{2}}-xy \frac{\p^{2}g(x,y)}{\p x\,\p y}+[c_{1}-(a+b_{1}+1)x]\frac{\p g(x,y)}{\p x}-b_{1}y\frac{\p g(x,y)}{\p y}-ab_{1}g(x,y)=0\,,\nn\\
&&y(1-y)\frac{\p^{2}g(x,y)}{\p y^{2}}-xy \frac{\p^{2}g(x,y)}{\p x\,\p y}+[c_{2}-(a+b_{2}+1)y]\frac{\p g(x,y)}{\p y}-b_{2}x\frac{\p g(x,y)}{\p x}-ab_{2}g(x,y)=0\,.\nn\\
\eea
The most general solution of these equations is 
\bea
g(x,y)&=&C_{1}\,F_{2}(a,b_{1},b_{2},c_{1},c_{2};x,y)+C_{2}\,x^{1-c_{1}}F_{2}(a-c_{1}+1,b_{1}-c_{1}+1,b_{2},2-c_{1},c_{2};x,y)\nn\\
&&+C_{3}\,y^{1-c_{2}}\,F_{2}(a-c_{2}+1,b_{1},b_{2}-c_{2}+1,c_{1},2-c_{2};x,y)\nn\\
&&+C_{4}\,x^{1-c_{1}}y^{1-c_{2}}\,F_{2}(a-c_{1}-c_{2}+2,b_{1}-c_{1}+1,b_{2}-c_{2}+1,2-c_{1},2-c_{2};x,y)
\eea
Here $F_{2}(a,b_{1},b_{2},c_{1},c_{2};x,y)$ is the Appell $F_{2}$ hypergeometric function, and for $|x|+|y|<1$, it has the following series representation
\be\label{F2Series}
F_{2}(a,b_{1},b_{2},c_{1},c_{2};x,y)=\sum_{m,n=0}^{\infty}\frac{(a)_{m+n}(b_{1})_{m}(b_{2})_{n}}{(c_{1})_{m}(c_{2})_{n}m!n!}x^{m}y^{n}\,.
\ee
In particular, note that we can partially sum the above series as
\be
F_{2}(a,b_{1},b_{2},c_{1},c_{2};u,v)=\sum_{m=0}^{\infty}\frac{(a)_{m}(b_{2})_{m}}{(c_{2})_{m}m!}\,v^{m}{}_{2}F_{1}(a+m,b_{1};c_{1};u)\,.
\ee
The Appell $F_{2}$ is relevant in our case because when we add the two equations in \eqref{AppellF2}, we get the differential equation~\eqref{MasterEq} provided we make the following identifications,
\bea
&&c_{1}=\frac{d}{2}+2-\Delta_{2},\quad c_{2}=\frac{d}{2}\,,\nn\\
&& a(b_{1}+b_{2})=(d+2-\frac{\Delta}{2})(\frac{d}{2}+\Delta_{1}-\frac{\Delta}{2}+1),\quad a+b_{1}+b_{2}=\frac{3d}{2}+\Delta_{1}-\Delta+3\,.
\eea
Thus, the Appell F$_{2}$ is a solution of the differential equation~\eqref{MasterEq}. However, it is not the complete solution as it is evident from the solution~\eqref{3PtFn2} that the most general solution is labelled by a priori unknown function of $u$. 
In the next section, we will obtain an explicit form of the 3-point function as a product of 2-point functions, thus providing an interpretation in terms of the Feynman diagram in momentum space.
%%%%%%%%%%%%%%%%%%%%%%%%%%%%%%%%%%%%%%%%%%%%%%%
\section{Explicit Fourier transform}
In this section, we obtain the momentum space correlation function by directly performing the Fourier transform of the position space correlations function~\eqref{FourierTrnsf}.  
We start with the 2-point function. The Fourier transform is
\bea
&&\int_{0}^{\infty}d\tau\int d^{d}x\,e^{-iE_{1}\tau}e^{-i\vec p_{1}\cdot x}\frac{C_{\mathcal O}}{t^{\Delta}}e^{-\frac{N_{1}\vec x^{2}}{2\tau}}=\frac{(2\pi)^{\frac{d}{2}}C_{\mathcal O}}{N_{1}^{\frac{d}{2}}}\Big(iE_{1}+\frac{\vec p_{1}^{2}}{2N_{1}}\Big)^{\Delta-\frac{d}{2}-1}\Gamma(1+\frac{d}{2}-\Delta)\,.
\eea
From the above expression, we note that the integral is not convergent when $\Delta-\frac{d}{2}-1=n$, where $n$ is a non-negative integer. We see this from the gamma function, which has a simple pole for $\Delta=\frac{d}{2}+1+n$. We have seen previously that the 2-point function in the momentum space is not well defined and requires regularization. The renormalized 2-point function is given in~\eqref{renorm2PtFn}.

\begin{figure}[htpb]
\begin{center}
\vspace{-2cm}
\centering
\includegraphics[width=6in]{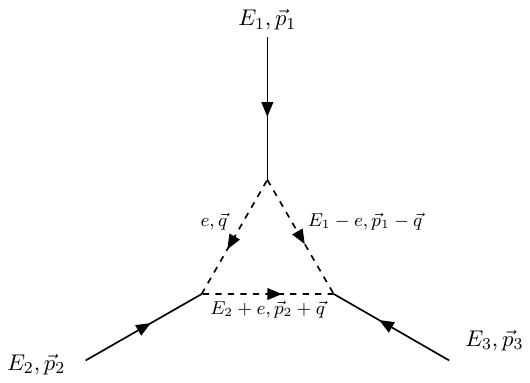}
\vspace{-11 cm}
\caption{Diagrammatic representation of 3-point function. Note that the particle number is conserved at every vertex. The internal lines $(e,\vec q), (E_{2}+e,\vec p_{2}+\vec q)$ and $(E_{1}-e,\vec p_{1}-\vec q)$ carry particle number $2\alpha, N_{2}+2\alpha$ and $N_{1}-2\alpha$, respectively. The diagram is drawn using the Tikz-Feynman package~\cite{Ellis:2016jkw}.}\label{DiagThreePtFn}
\end{center}
\end{figure}

Next, we discuss the Fourier transform of the 3-point function. We start with the Fourier transform of the spatial part of the correlation function
\be
I(p_{i})=\int\prod_{i=1}^{3}d^{d}x_{i}\,e^{-i\vec p_{i}\cdot \vec x_{i}}e^{-\frac{N_{1}}{2}\frac{\vec x_{13}^{2}}{\tau_{13}}-\frac{N_{2}}{2}\frac{\vec x_{23}^{2}}{\tau_{23}}}F(v_{123})\,.
\ee
For the later purposes, it will be convenient to choose $F(v_{123})=e^{-\alpha v_{123}}=e^{-\alpha(\frac{\vec x_{23}^{2}}{\tau_{23}}+\frac{\vec x_{12}^{2}}{\tau_{12}}-\frac{\vec x_{13}^{2}}{\tau_{13}})}$.
Then we have the integration
\be
I(p_{i},\alpha)=\int\prod_{i=1}^{3}d^{d}x_{i}\,e^{-i\vec p_{i}\cdot \vec x_{i}}e^{-\frac{N_{1}}{2}\frac{\vec x_{13}^{2}}{\tau_{13}}-\frac{N_{2}}{2}\frac{\vec x_{23}^{2}}{\tau_{23}}}e^{-\alpha(\frac{\vec x_{23}^{2}}{\tau_{23}}+\frac{\vec x_{12}^{2}}{\tau_{12}}-\frac{\vec x_{13}^{2}}{\tau_{13}})}\,.
\ee
We can express the above Fourier transform as the convolution of the product of the Fourier transform of 2-point functions. In order to do this, we express the integral as
\bea\label{Fourier3pt.1}
I(p_{i},\alpha)&=&\int\prod_{i=1}^{3}d^{d}x_{i}d^{d}y_{i}\,e^{-ip_{i}x_{i}}\,\prod_{i=1}^{3}\delta^{d}(x_{i}-y_{i})\,e^{-\frac{N_{1}}{2}\frac{x_{13}^{2}}{t_{13}}-\frac{N_{2}}{2}\frac{x_{23}^{2}}{t_{23}}}e^{\alpha(\frac{x_{23}^{2}}{t_{23}}+\frac{x_{12}^{2}}{t_{12}}-\frac{x_{13}^{2}}{t_{13}})}\,,\nn\\
&=&\int\prod_{i=1}^{3}\frac{d^{d}q_{i}}{(2\pi)^{d}}\int\prod_{i=1}^{3}d^{d}x_{i}d^{d}y_{i}\,e^{-ip_{i}x_{i}+iq_{i}(x_{i}-y_{i})}\,e^{-\frac{\tilde N_{1}}{2}\frac{\tilde x_{13}^{2}}{t_{13}}-\frac{\tilde N_{2}}{2}\frac{\tilde x_{23}^{2}}{t_{23}}-\frac{\tilde N_{3}}{2}\frac{\tilde x_{12}^{2}}{t_{12}}}\,.
\eea
Here,
\be
\tilde N_{1}=N_{1}-2\alpha,\quad \tilde N_{2}=N_{2}+2\alpha\,\quad \tilde N_{3}=2\alpha,\quad \vec{\tilde x}_{13}=\vec x_{1}-\vec y_{3},\,\,\vec{\tilde x}_{23}=\vec y_{2}-\vec x_{3},\,\,\vec{\tilde x}_{12}=\vec y_{1}-\vec x_{2}\,.
\ee
Then, the integral~\eqref{Fourier3pt.1} can be written as
\be
I(p_{i},\alpha)=\int\prod_{i=1}^{3}\frac{d^{d}q_{i}}{(2\pi)^{d}} f_{13}(\vec p_{1}-\vec q_{1},\vec q_{3})f_{23}(\vec q_{2},\vec p_{3}-\vec q_{3})f_{12}(\vec q_{1},\vec p_{2}-\vec q_{2})\,,
\ee
where
\be
f_{ab}(\vec p_{a},\vec q_{b})=\int d^{d}z_{a}\,d^{d}z_{b}\,e^{-i(\vec p_{a}\cdot \vec z_{a}+\vec q_{b}\cdot \vec z_{b})}e^{-\frac{\tilde N_{a}(\vec z_{a}-\vec z_{b})^{2}}{2\tau_{ab}}}=(2\pi)^{d}\delta^{d}(p_{a}+q_{b})\Big(\frac{2\pi \tau_{ab}}{\tilde N_{a}}\Big)^{\frac{d}{2}}e^{-\frac{\tau_{ab}}{2\tilde N_{a}}\vec p_{a}^{2}}\,.
\ee
Thus, using all the delta functions we obtain
\be
I(p_{i},\alpha)=(2\pi)^{d}\delta^{d}(p_{1}+p_{2}+p_{3})\Big(\frac{(2\pi)^{3}\tau_{13}\tau_{23}\tau_{12}}{\tilde N_{1}\tilde N_{2}\tilde N_{3}}\Big)^{\frac{d}{2}}\int\frac{d^{d}q}{(2\pi)^{d}}e^{-\frac{\tau_{13}}{2\tilde N_{1}}(\vec p_{1}-\vec q)^{2}-\frac{\tau_{23}}{2\tilde N_{2}}(\vec p_{2}+\vec q)^{2}-\frac{\tau_{12}}{2\tilde N_{3}}\vec q^{2}}\,.
\ee
Now, let us calculate the rest of the integrals
\be
J(E_{i},p_{i},\alpha)=\int_{-\infty}^{\infty}\prod_{i=1}^{3}d\tau_{i}\,e^{-iE_{i}\tau_{i}}\tau_{12}^{\Delta_{12}}\tau_{13}^{\Delta_{13}}\tau_{23}^{\Delta_{23}}I(p_{i},\alpha)\,\Theta(\tau_{1}-\tau_{2})\Theta(\tau_{2}-\tau_{3})\Theta(\tau_{1}-\tau_{3})\,.
\ee
Proceeding in the same manner as we did with momentum integration, we obtain
\be
J(E_{i},p_{i},\alpha)
=(2\pi)^{d}\delta^{d}(p_{1}+p_{2}+p_{3})\Big(\frac{(2\pi)^{3}}{\tilde N_{1}\tilde N_{2}\tilde N_{3}}\Big)^{\frac{d}{2}}\int\prod_{i=1}^{3}\frac{de_{i}}{(2\pi)}g(e_{1},E_{2}-e_{2})g(e_{2},E_{3}-e_{3})g(E_{1}-e_{1},e_{3})\,,
\ee
where
\be
g(e,E)=\int dt\,dT\,e^{-i(Et+eT)-\frac{(t-T)p^{2}}{2m}}(t-T)^{\delta}\Theta(t-T)=(2\pi)\delta(e+E)\,(iE+\frac{p^{2}}{2m})^{-1-\delta}\Gamma(1+\delta)\,.
\ee
Using the above expression, our final integration is
\bea\label{3PtFn3}
J(E_{i},p_{i},\alpha)&=&(2\pi)^{d+1}\delta^{d}(p_{1}+p_{2}+p_{3})\delta^{d}(E_{1}+E_{2}+E_{3})\Big(\frac{(2\pi)^{3}}{\tilde N_{1}\tilde N_{2}\tilde N_{3}}\Big)^{\frac{d}{2}}\Gamma(1+\Delta_{12}+\frac{d}{2})\times\nn\\
&&\times\Gamma(1+\Delta_{13}+\frac{d}{2})\Gamma(1+\Delta_{23}+\frac{d}{2})\int\frac{de\,d^{d}q}{(2\pi)^{d+1}}\Big(i(E_{1}-e)+\frac{(p_{1}-q)^{2}}{2\tilde N_{1}}\Big)^{-1-\Delta_{13}-\frac{d}{2}}\times\nn\\
&&\times\Big(i(E_{2}+e)+\frac{(p_{2}+q)^{2}}{2\tilde N_{2}}\Big)^{-1-\Delta_{23}-\frac{d}{2}} \Big(ie+\frac{q^{2}}{2\tilde N_{3}}\Big)^{-1-\Delta_{12}-\frac{d}{2}}\,,\nn\\
&=&(2\pi)^{d+1}\delta^{d}(p_{1}+p_{2}+p_{3})\delta^{d}(E_{1}+E_{2}+E_{3})<<J(E_{i},p_{i},\alpha)>>\,.
\eea
where $<<J(E_{i},p_{i},\alpha)>>$  can be identified with $<<O(E_{1},\vec p_{1})O(E_{2},\vec p_{2})O(E_{3},\vec p_{3})>>$ for the above choice of the function $F(v_{123})$.
We see that the 3-point function has the structure of one loop computation and has a diagrammatic representation as shown in the Figure~\ref{DiagThreePtFn}. Furthermore, the parameter $\alpha$ used in the Fourier transform behaves like the particle number. The form of the function $F(v_{123})$ is a theory dependent; nevertheless, given the Fourier transform for $F(v_{123})=e^{-\alpha v_{123}}$, and assuming that the integral~\eqref{3PtFn3} exists, we can write the general 3-point function as the inverse Laplace transform, i.e
\be
<<O(E_{1},\vec p_{1})O(E_{2},\vec p_{2})O(E_{3},\vec p_{3})>>=\frac{1}{2\pi i}\int_{\g-i\infty}^{\g+i\infty}d\alpha\,\rho(\alpha)J(E_{i},p_{i},\alpha)\,,
\ee
for a suitable choice of $\g$. Here, $\rho(\alpha)$ is a theory dependent function which can be identified as the spectral function. The function $\rho(\alpha)$ contains the information of the particle number of operators running inside the loop.

In a similar fashion, we can also express the 4-point function of scalar operators in momentum space. The momentum space 4-point function has been obtained previously~\cite{Mehen:1999nd}; however, here, we will express it in terms of the product of the 2-point function, which is reminiscent of Feynman diagram computations in the momentum space.

\begin{figure}[htpb]
\begin{center}
\vspace{-1cm}
\centering
\includegraphics[width=3in]{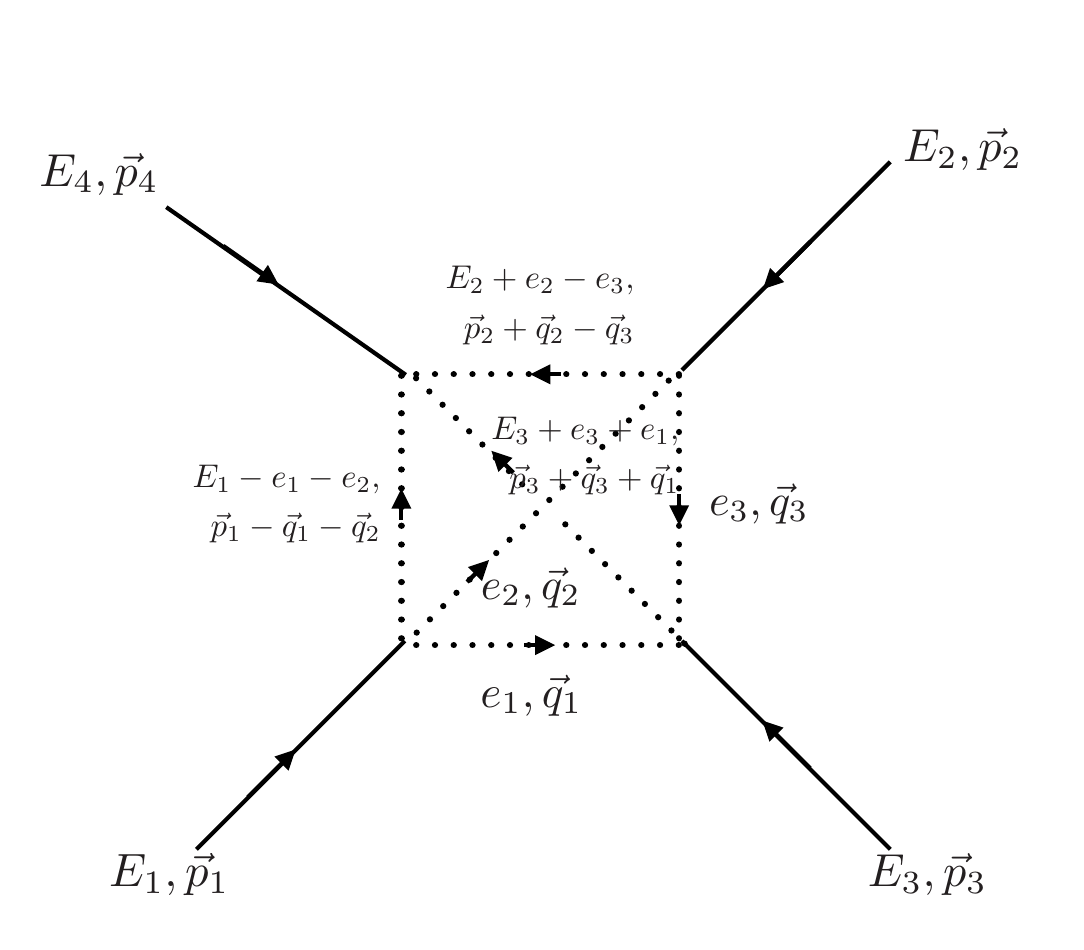}
%\vspace{-1 cm}
\caption{Diagrammatic representation of 4-point function. Note that the particle number is conserved at every vertex.}\label{DiagFourPtFn}
\end{center}
\end{figure}

The position space correlation function depends on an arbitrary function of four cross ratios, i.e. $h(v_{124},v_{134},v_{234},\frac{\tau_{12}\tau_{34}}{\tau_{13}\tau_{24}})$. Choosing
\be
h(v_{124},v_{134},v_{234},\frac{\tau_{12}\tau_{34}}{\tau_{13}\tau_{24}})=e^{-\alpha_{1}v_{134}-\alpha_{2}v_{124}-\alpha_{3}v_{234}}\Big(\frac{\tau_{12}\tau_{34}}{\tau_{13}\tau_{24}}\Big)^{\delta}\,,
\ee
and performing the Fourier transform, we obtain
\be
<<O_{1}(E_{1},\vec p_{1})O_{1}(E_{2},\vec p_{2})O_{1}(E_{3},\vec p_{3})O_{1}(E_{4},\vec p_{4})>>_{\{\alpha_{i}\},\delta}=H(\{\alpha_{i}\},\delta)\,.
\ee
Here
\bea\label{4PtIntegral}
H(\{\alpha_{i}\},\delta)&=&\int \prod_{i=1}^{3}\frac{de_{i}\,d^{d}q_{i}}{(2\pi)^{d+1}}\Big(ie_{1}+\frac{q_{1}^{2}}{2\tilde N_{1}}\Big)^{-1-\delta_{1}}\Big(ie_{2}+\frac{q_{2}^{2}}{2\tilde N_{2}}\Big)^{-1-\delta_{2}}\Big(ie_{3}+\frac{q_{3}^{2}}{2\tilde N_{3}}\Big)^{-1-\delta_{3}}\nn\\
&&\Big(i(E_{1}-e_{1}-e_{2})+\frac{(\vec p_{1}-\vec q_{1}-\vec q_{2})^{2}}{2\tilde N_{4}}\Big)^{-1-\delta_{4}}\times\Big(i(E_{2}+e_{2}-e_{3})+\frac{(\vec p_{2}+\vec q_{2}-\vec q_{3})^{2}}{2\tilde N_{5}}\Big)^{-1-\delta_{5}}\nn\\
&&\times\Big(i(E_{3}+e_{1}+e_{3})+\frac{(\vec p_{3}+\vec q_{1}+\vec q_{3})^{2}}{2\tilde N_{6}}\Big)^{-1-\delta_{6}}\,.
\eea
The above expression has the structure of a 3-loop Feynman diagram as shown in the figure~\ref{DiagFourPtFn}. Interestingly, this is the same structure that was obtained for the 4-point function in the relativistic CFT.
The internal lines have the definite particle number given as 
\bea
\tilde N_{1}=2\alpha_{1},\quad \tilde N_{2}=2\alpha_{2},\quad \tilde N_{3}=2\alpha_{3},\quad \tilde N_{4}=N_{1}-2\alpha_{1}-2\alpha_{2},\quad \tilde N_{5}=N_{2}+2\alpha_{2}-2\alpha_{3},\quad \tilde N_{6}=N_{3}+2\alpha_{1}+2\alpha_{3}
\eea
and the scaling dimensions that are determined by
\bea
&&\delta_{1}=\frac{\Delta}{6}-\frac{\Delta_{1}+\Delta_{2}}{2}+\frac{d}{2}+\delta,\quad \delta_{2}=\frac{\Delta}{6}-\frac{\Delta_{1}+\Delta_{3}}{2}+\frac{d}{2}-\delta,\quad \delta_{3}=\frac{\Delta}{6}-\frac{\Delta_{1}+\Delta_{4}}{2}+\frac{d}{2}\,,\nn\\
&&\delta_{4}=\frac{\Delta}{6}-\frac{\Delta_{2}+\Delta_{3}}{2}+\frac{d}{2},\quad \delta_{5}=\frac{\Delta}{6}-\frac{\Delta_{2}+\Delta_{4}}{2}+\frac{d}{2}-\delta,\quad \delta_{6}=\frac{\Delta}{6}-\frac{\Delta_{3}+\Delta_{4}}{2}+\frac{d}{2}+\delta\,,\nn\\
\eea
where $\Delta=\sum_{a=1}^{6}\Delta_{a}$.

Note that at every vertex, the particle number is conserved. Assuming that the integral~\eqref{4PtIntegral} is convergent, the general expression for the 4-point function in the momentum space can be given by
\be
<<O_{1}(E_{1},\vec p_{1})O_{1}(E_{2},\vec p_{2})O_{1}(E_{3},\vec p_{3})O_{1}(E_{4},\vec p_{4})>>=\int_{\g_{1}-i\infty}^{\g_{1}+i\infty}\int_{\g_{2}-i\infty}^{\g_{2}+i\infty}\int_{\g_{3}-i\infty}^{\g_{3}+i\infty}\prod_{i=1}^{3}\frac{d\alpha_{i}}{2\pi i}\int_{\g_{4}-i\infty}^{\g_{4}+i\infty}\frac{d\delta}{2\pi i}\rho(\{\alpha_{i}\},\delta)H(\{\alpha_{i}\},\delta)\,.
\ee
Here $\g_{i}$ and $\rho(\{\alpha_{i}\},\delta)$ are suitable constants and spectral function, respectively.

The integrals~\eqref{3PtFn3} and~\eqref{4PtIntegral} may diverge for certain scaling dimensions. Similar to the 2-point function discussed previously, it needs regularization and renormalization.
%%%%%%%%%%%%%%%%%%%%%%%%%%%%%%%%%%%%%%%%%%%%%%%%%%%%
\section{Generalization to NRCFT in the presence of boundary}
The above analysis can be generalized to NRCFT's in the presence of a boundary~\cite{Gupta:2022azd}. In this case, the correlation functions of bulk operators satisfy the Ward identities for those generators that preserve the boundary. The Ward identities for scalar operators, in the case of a planar boundary, are
\bea\label{BoundaryWardidSpatial}
&&\sum_{a=1}^{n}\Big(2\tau_{a}\frac{\p}{\p \tau_{a}}+x_{ai}\frac{\p}{\p x_{ai}}+y_{a}\frac{\p}{\p y_{a}}+\Delta_{a}\Big)<{\mathcal T}O_{1}(\tau_{1},\vec x_{1},y_{1})......O_{n}(\tau_{n},\vec x_{n},y_{n})>=0\,,\nn\\
&&\sum_{a=1}^{n}\Big(\tau_{a}^{2}\frac{\p}{\p \tau_{a}}+\tau_{a}x_{ai}\frac{\p}{\p x_{ai}}+\tau_{a}y_{a}\frac{\p}{\p y_{a}}+\tau_{a}\Delta_{a}+\frac{N_{a}}{2}x_{a}^{2}+\frac{N_{a}}{2}y_{a}^{2}\Big)<{\mathcal T}O_{1}(\tau_{1},\vec x_{1},y_{1})......O_{n}(\tau_{n},\vec x_{n},y_{n})>=0\,,\nn\\
&&\sum_{a=1}^{n}\Big(\tau_{a}\frac{\p}{\p x_{ai}}+N_{a}x_{ai}\Big)<{\mathcal T}O_{1}(\tau_{1},\vec x_{1},y_{1})......O_{n}(\tau_{n},\vec x_{n},y_{n})>=0\,.
\eea
Here, $\vec x$ and $y$ represent the boundary and normal coordinates, respectively. We have assumed that the boundary is at $y=0$. Next, we will express the above Ward identities in the momentum space.

The particle number conservation and time and boundary space translation invariance implies that
\bea
&&<O_{1}(E_{1},\vec p_{1},y_{1}).....O_{n}(E_{n},\vec p_{n},y_{n})>\nn\\
\quad\quad\quad&=&(2\pi)^{d+1}\delta_{N_{1}+..+N_{n},0}\delta(E_{1}+...+E_{n})\delta^{d}(\vec p_{1}+...+\vec p_{n})<<O_{1}(E_{1},\vec p_{1},y_{1})......O_{n}(E_{n},\vec p_{n},y_{n})>>\,.\nn\\
\eea
Here $d>1$ is the spatial dimension of the boundary. Thus, momentum space correlation function will depend on $(n-1)$-energies $E$ and momentum $\vec p$ and $n$-normal coordinates $y_{a}$, for $a=1,..,n$. 

The dilatation Ward identity in the momentum space becomes
\be
\Big[\sum_{a=1}^{n}(\Delta_{a}+y_{a}\frac{\p}{\p y_{a}})-(n-1)(d+2)-\sum_{a=1}^{n-1}\Big(p_{a,i}\frac{\p}{\p p_{a,i}}+2E_{a}\frac{\p}{\p E_{a}}\Big)\Big]<<O_{1}(E_{1},\vec p_{1})......O_{n}(E_{n},\vec p_{n})>>=0\,.
\ee
The expansion Ward identity is
\bea
\sum_{a=1}^{n-1}\Big(-i(d+2-\Delta_{a})\frac{\p}{\p E_{a}}-iE_{a}\frac{\p^{2}}{\p E_{a}^{2}}-ip_{ai}\frac{\p^{2}}{\p E_{a}\p p_{ai}}-\frac{N_{a}}{2}\frac{\p^{2}}{\p p_{ai}\p p_{ai}}+iy_{a}\frac{\p^{2}}{\p E_{a}\p y_{a}}\Big)<<O_{1}(E_{1},\vec p_{1})......O_{n}(E_{n},\vec p_{n})>>\nn\\
 \qquad\qquad \qquad\qquad +\sum_{a=1}^{n}(\frac{N_{a}}{2}y_{a}^{2})<<O_{1}(E_{1},\vec p_{1})......O_{n}(E_{n},\vec p_{n})>>=0\,.
\eea
Finally, the boost Ward identity is
\be\label{BoostId-bdy}
\sum_{a=1}^{n-1}\Big(-p_{ai}\frac{\p}{\p E_{a}}+iN_{a}\frac{\p}{\p p_{ai}}\Big)<<O_{1}(E_{1},\vec p_{1},y_{1})......O_{n}(E_{n},\vec p_{n},y_{n})>>=0\,.
\ee
The solution to the position space Ward identities have been discussed in~\cite{Gupta:2022azd}. In this case, the one point function can have non-zero expectation value provided the operator has vanishing particle number. The two point function is fixed upto a undetermined function of the cross ration $\xi=\frac{y_{1}y_{2}}{t_{12}}$. Motivated by the position space analysis, we look for the solution to the Ward identities for the 2-point function.

The boost Ward identity implies that the solution depends on energy and momentum through $\omega=iE+\frac{\vec p^{2}}{2m}$, i.e.
\be
<<O_{1}(E,\vec p,y_{1})O_{2}(-E,-\vec p,y_{2})>>=f(\omega,y_{1},y_{2})\,.
\ee
Using the dilatation Ward identity, we can write
\be
f(\omega,y_{1},y_{2})=\omega^{c_{0}}g(\eta_{1},\eta_{2})\,,
\ee
where $\eta_{1,2}=y_{1,2}^{2}\omega$ and $c_{0}=\Delta-\frac{d}{2}-1$. Substituting the above in the expansion Ward identity, we obtain the following differential equation
\be\label{ModExpansionWardId}
\eta_{1}^{2}\frac{\p^{2}}{\p\eta_{1}^{2}}g(\eta_{1},\eta_{2})+(1+c_{0})\eta_{1}\frac{\p}{\p\eta_{1}}g(\eta_{1},\eta_{2})-\frac{N_{1}}{2}\eta_{1}g(\eta_{1},\eta_{2})-\Big(\eta_{2}^{2}\frac{\p^{2}}{\p\eta_{2}^{2}}g(\eta_{1},\eta_{2})+(1+c_{0})\eta_{2}\frac{\p}{\p\eta_{2}}g(\eta_{1},\eta_{2})-\frac{N_{1}}{2}\eta_{2}g(\eta_{1},\eta_{2})\Big)=0\,.
\ee
The general solution of the above can be obtained using the separation of variables in which case we need to solve two identical equations of the form
\be
\eta_{1}^{2}\frac{\p^{2}}{\p\eta_{1}^{2}}g_{1}(\eta_{1})+(1+c_{0})\eta_{1}\frac{\p}{\p\eta_{1}}g_{1}(\eta_{1})-(b+\frac{N_{1}}{2}\eta_{1})g_{1}(\eta_{1})=0\,,
\ee
where $b$ is a complex constant. For a given constant $b$, the general solution is a linear combination of the modified Bessel function of the first kind, i.e.
\be
g_{1}(\eta_{1};b)=\frac{1}{\eta_{1}^{\frac{c_{0}}{2}}}\Big(a_{1}I_{i\nu}(\sqrt{2N_{1}\eta_{1}})+a_{2}I_{-i\nu}(\sqrt{2N_{1}\eta_{1}})\Big)\,.
\ee
Here $i\nu=\sqrt{4b+c_{0}^{2}}$. Thus, the complete solution of~\eqref{ModExpansionWardId} for a given constant $b$ is
\be\label{GeneralSoln.bdy}
g(\eta_{1},\eta_{2};b)=\frac{1}{(\eta_{1}\eta_{2})^{\frac{c_{0}}{2}}}\Big(a_{1}I_{i\nu}(\sqrt{2N_{1}\eta_{1}})+a_{2}I_{-i\nu}(\sqrt{2N_{1}\eta_{1}}\Big)\Big(\tilde a_{1}I_{i\nu}(\sqrt{2N_{1}\eta_{2}})+\tilde a_{2}I_{-i\nu}(\sqrt{2N_{1}\eta_{2}}\Big)\,.
\ee
The solution contains constants $a_{i},\tilde a_{i}$ and $b$. The constants $a_{i}$ and $\tilde a_{i}$ are fixed by looking at the asymptotic of the above solution. We note that for a given $\eta_{2}$, the general solution~\eqref{GeneralSoln.bdy} diverges for $\eta_{1}\rightarrow\infty$, unless $a_{1}=-a_{2}$. This follows from the asymptotic of the modified Bessel function
\be
I_{i\nu}(z)\sim \frac{e^{z}}{\sqrt{2\pi z}}\,,\quad\text{for}\quad|z|>>1\,.
\ee
Assuming that $g(\eta_{1},\eta_{2};b)$ should be symmetric in $\eta_{1}$ and $\eta_{2}$, we have the solution in terms of the product of the modified Bessel function of the second kind i.e.
\be
g(\eta_{1},\eta_{2};b)=\frac{A}{(\eta_{1}\eta_{2})^{\frac{c_{0}}{2}}}K_{i\nu}(\sqrt{2N_{1}\eta_{1}})K_{i\nu}(\sqrt{2N_{1}\eta_{2}})\,.
\ee 
Finally, we can write the general solution of the conformal Ward identity to be
\be\label{FinalSol2Ptb}
<<O_{1}(E,\vec p,y_{1})O_{2}(-E,-\vec p,y_{2})>>=\frac{\omega^{c_{0}}}{(\eta_{1}\eta_{2})^{\frac{c_{0}}{2}}}\int_{\Gamma}d\nu\,\rho(\nu)K_{i\nu}(\sqrt{2N_{1}\eta_{1}})K_{i\nu}(\sqrt{2N_{1}\eta_{2}})\,,
\ee
for some choice of the contour $\Gamma$ and the spectral function $\rho(\nu)$.

It would be illustrative to compare the above solution~\eqref{FinalSol2Ptb} with the direct Fourier transform of the position space two point function. In the position space, the two point function is determined up to a function of a cross ratio $\xi=\frac{y_{1}y_{2}}{\tau_{12}}$ and is given by~\cite{Gupta:2022azd}
\be
<\mathcal T\mathcal O_{1}(\tau_{1}, \vec x_{1},y_{1})\mathcal O_{2}(\tau_{2}, \vec x_{2},y_{2})>=\frac{\mathcal A}{t_{12}^{\Delta}}e^{-\frac{N_{1}}{2\tau_{12}}(\vec x_{12}^{2}+y_{1}^{2}+y_{2}^{2})}G(\xi)\,,
\ee
where $\mathcal A$ is a constant. The Fourier transform of the above for a choice $G(\xi)=\xi^{\chi}$ is
\be\label{2PtbFourier}
\tilde{\mathcal A} \,\Big(\frac{\omega}{\sqrt{\eta_{1}\eta_{2}}}\Big)^{c_{0}}\Big(\frac{N\eta_{1}\eta_{2}}{2(\eta_{1}+\eta_{2})}\Big)^{\frac{\chi+c_{0}}{2}}K_{\chi+c_{0}}(\sqrt{2N_{1}(\eta_{1}+\eta_{2}})\,,
\ee
where $\tilde{\mathcal A}$ is a constant and we have hidden the delta function dependence. Next, we would like to find the choice of the contour $\Gamma$ and the function $\rho(\nu)$ in~\eqref{FinalSol2Ptb} that reproduces the above. Using the integral identity~\cite{Bateman,Gradshteyn:1943cpj}
\be\label{IntegralId}
\int_{0}^{\infty}d\nu\,\nu\sinh(\pi\nu)\Gamma(\lambda+\frac{i}{2}\nu)\Gamma(\lambda-\frac{i}{2}\nu)K_{i\nu}(x)K_{i\nu}(y)=2\pi^{2}\Big(\frac{xy}{2\sqrt{x^{2}+y^{2}}}\Big)^{2\lambda}K_{2\lambda}(\sqrt{x^{2}+y^{2}}),
\ee
which is true for $\text{Re}\lambda>0$, we find that the momentum space correlator~\eqref{FinalSol2Ptb} precisely agrees with~\eqref{2PtbFourier} with 
\be
\rho(\nu)=\frac{\tilde{\mathcal A}}{2\pi^{2}}\,\nu\sinh(\pi\nu)\Gamma(\frac{\chi+c_{0}}{2}+\frac{i}{2}\nu)\Gamma(\frac{\chi+c_{0}}{2}-\frac{i}{2}\nu)\,,
\ee
for $\text{Re}(\chi+c_{0})>0$. This determines the choice of $\Gamma$ and the function $\rho(\nu)$.

%%%%%%%%%%%%%%%%%%%%%%%%%%%%%%%%%%%%%%%%%%%%%%%%%%%
\section*{Acknowledgments}
The work of R Gupta is supported by SERB MATRICS grant MTR/2022/000291 and CRG/2023/001388. 
Meenu would like to thank the Council of Scientific and Industrial Research (CSIR), Government of India, for the financial support through a research fellowship (Award No.09/1005(0038)/2020-EMR-I).
%%%%%%%%%%%%%%%%%%%%%%%%%%%%%%%%%%%%%%%%%%%%%%%%%%%
\providecommand{\href}[2]{#2}\begingroup\raggedright\endgroup

\end{document}